\input harvmac.tex
\input epsf
\noblackbox
\def\figin{\epsfcheck\figin}\def\figins{\epsfcheck\figins}
\def\epsfcheck{\ifx\epsfbox\UnDeFiNeD
\message{(NO epsf.tex, FIGURES WILL BE IGNORED)}
\gdef\figin##1{\vskip2in}\gdef\figins##1{\hskip.5in}
\else\message{(FIGURES WILL BE INCLUDED)}%
\gdef\figin##1{##1}\gdef\figins##1{##1}\fi}
\def\DefWarn#1{}
\def\figinsert{\goodbreak\midinsert}
\def\ifig#1#2#3{\DefWarn#1\xdef#1{fig.~\the\figno}
\writedef{#1\leftbracket fig.\noexpand~\the\figno}%
\figinsert\figin{\centerline{#3}}\medskip\centerline{\vbox{\baselineskip12pt
\advance\hsize by -1truein\noindent\footnotefont{\bf Fig.~\the\figno } \it#2}}
\bigskip\endinsert\global\advance\figno by1}


\def\encadremath#1{\vbox{\hrule\hbox{\vrule\kern8pt\vbox{\kern8pt
 \hbox{$\displaystyle #1$}\kern8pt}
 \kern8pt\vrule}\hrule}}
 %
 %
 

 \font\cmss=cmss10
 \font\cmsss=cmss10 at 7pt
 \def\rlx{\relax\leavevmode}
 \def\inbar{\vrule height1.5ex width.4pt depth0pt}
 \def\IC{\relax\,\hbox{$\inbar\kern-.3em{\rm C}$}}
 \def\IN{\relax{\rm I\kern-.18em N}}
 \def\IP{\relax{\rm I\kern-.18em P}}

\def\ZZ{\rlx\leavevmode\ifmmode\mathchoice{\hbox{\cmss Z\kern-.4em Z}}
  {\hbox{\cmss Z\kern-.4em Z}}{\lower.9pt\hbox{\cmsss Z\kern-.36em Z}}
  {\lower1.2pt\hbox{\cmsss Z\kern-.36em Z}}\else{\cmss Z\kern-.4em Z}\fi}
 \def\IZ{\relax\ifmmode\mathchoice
 {\hbox{\cmss Z\kern-.4em Z}}{\hbox{\cmss Z\kern-.4em Z}}
 {\lower.9pt\hbox{\cmsss Z\kern-.4em Z}}
 {\lower1.2pt\hbox{\cmsss Z\kern-.4em Z}}\else{\cmss Z\kern-.4em Z}\fi}
 \def\IZ{\relax\ifmmode\mathchoice
 {\hbox{\cmss Z\kern-.4em Z}}{\hbox{\cmss Z\kern-.4em Z}}
 {\lower.9pt\hbox{\cmsss Z\kern-.4em Z}}
 {\lower1.2pt\hbox{\cmsss Z\kern-.4em Z}}\else{\cmss Z\kern-.4em Z}\fi}

 \def\narrowplus{\kern -.04truein + \kern -.03truein}
 \def\narrowminus{- \kern -.04truein}
 \def\narrowminussub{\kern -.02truein - \kern -.01truein}

 \def\frac#1#2{{#1\over #2}}

 \def\sym#1{{{\rm SYM}} _{#1 +1}}
 
 \def\IZ{\relax\ifmmode\mathchoice
 {\hbox{\cmss Z\kern-.4em Z}}{\hbox{\cmss Z\kern-.4em Z}}
 {\lower.9pt\hbox{\cmsss Z\kern-.4em Z}}
 {\lower1.2pt\hbox{\cmsss Z\kern-.4em Z}}\else{\cmss Z\kern-.4em Z}\fi}
 \def\IB{\relax{\rm I\kern-.18em B}}
 \def\IC{{\relax\hbox{$\inbar\kern-.3em{\rm C}$}}}
 \def\Ic{{\relax\hbox{$\inbar\kern-.22em{\rm c}$}}}
 \def\ID{\relax{\rm I\kern-.18em D}}
 \def\IE{\relax{\rm I\kern-.18em E}}
 \def\IF{\relax{\rm I\kern-.18em F}}
 \def\IG{\relax\hbox{$\inbar\kern-.3em{\rm G}$}}
 \def\IGa{\relax\hbox{${\rm I}\kern-.18em\Gamma$}}
 \def\IH{\relax{\rm I\kern-.18em H}}
 \def\II{\relax{\rm I\kern-.18em I}}
 \def\IK{\relax{\rm I\kern-.18em K}}
 \def\IP{\relax{\rm I\kern-.18em P}}

 \font\cmss=cmss10 \font\cmsss=cmss10 at 7pt
 \def\IR{\relax{\rm I\kern-.18em R}}

 %

 %
 %
 \def\eqnn#1{\xdef
#1{(\secsym\the\meqno)}\writedef{#1\leftbracket#1}%
 \global\advance\meqno by1\wrlabeL#1}
 \def\eqna#1{\xdef
#1##1{\hbox{$(\secsym\the\meqno##1)$}}

\writedef{#1\numbersign1\leftbracket#1{\numbersign1}}%
 \global\advance\meqno by1\wrlabeL{#1$\{\}$}}
 \def\eqn#1#2{\xdef
#1{(\secsym\the\meqno)}\writedef{#1\leftbracket#1}%
 \global\advance\meqno by1$$#2\eqno#1\eqlabeL#1$$}

\newdimen\tableauside\tableauside=1.0ex
\newdimen\tableaurule\tableaurule=0.4pt
\newdimen\tableaustep
\def\phantomhrule#1{\hbox{\vbox to0pt{\hrule height\tableaurule width#1\vss}}}
\def\phantomvrule#1{\vbox{\hbox to0pt{\vrule width\tableaurule height#1\hss}}}
\def\sqr{\vbox{%
  \phantomhrule\tableaustep
  \hbox{\phantomvrule\tableaustep\kern\tableaustep\phantomvrule\tableaustep}%
  \hbox{\vbox{\phantomhrule\tableauside}\kern-\tableaurule}}}
\def\squares#1{\hbox{\count0=#1\noindent\loop\sqr
  \advance\count0 by-1 \ifnum\count0>0\repeat}}
\def\tableau#1{\vcenter{\offinterlineskip
  \tableaustep=\tableauside\advance\tableaustep by-\tableaurule
  \kern\normallineskip\hbox
    {\kern\normallineskip\vbox
      {\gettableau#1 0 }%
     \kern\normallineskip\kern\tableaurule}%
  \kern\normallineskip\kern\tableaurule}}
\def\gettableau#1 {\ifnum#1=0\let\next=\null\else
  \squares{#1}\let\next=\gettableau\fi\next}

\tableauside=1.0ex
\tableaurule=0.4pt

\def\IE{\relax{\rm I\kern-.18em E}}
\def\IP{\relax{\rm I\kern-.18em P}}

\lref\gvmone{R. Gopakumar and C. Vafa, ``M-theory and topological
strings, I," hep-th/9809187. }
\lref\gz{T. Graber and E. Zaslow, ``Open-string Gromov-Witten invariants:
calculations and a mirror `theorem','' hep-th/0109075.}
\lref\wittcs{E. Witten, ``Chern-Simons gauge theory as
a string theory,'' hep-th/9207094, in {\it The Floer memorial volume},
H. Hofer, C.H. Taubes, A. Weinstein and E. Zehner, eds.,
Birkh\"auser 1995, p. 637.}
\lref\gv{R. Gopakumar and C. Vafa, ``On the gauge theory/geometry
correspondence," hep-th/9811131, Adv. Theor. Math. Phys. {\bf 3} (1999)
1415.}
\lref\lm{J.M.F. Labastida and M. Mari\~no, ``Polynomial invariants
for torus knots and topological strings,''  hep-th/0004196,
Commun. Math. Phys. {\bf 217} (2001) 423.}
\lref\gvtwo{R. Gopakumar and C. Vafa, ``M-theory and topological
strings, II," hep-th/9812127. }
\lref\ov{H. Ooguri and C. Vafa, ``Knot invariants and topological
strings," hep-th/9912123, Nucl. Phys. {\bf B 577} (2000) 419.}
\lref\jones{E. Witten, ``Quantum field theory and the Jones polynomial,"
Commun. Math. Phys. {\bf 121} (1989) 351.}
\lref\lmv{J.M.F. Labastida, M. Mari\~no and C. Vafa, ``Knots, links and
branes at large $N$,'' hep-th/0010102, JHEP {\bf 0011} (2000) 007.}
\lref\dfg{E. Diaconescu, B. Florea and A. Grassi,
``Geometric transitions and open string instantons,'' hep-th/0205234.}
\lref\mv{M. Mari\~no and C. Vafa, ``Framed knots at large $N$,''
hep-th/0108064.}
\lref\macdonald{I.G. Macdonald, {\it Symmetric functions and Hall
polynomials}, 2nd edition, Oxford University Press, 1995.}
\lref\ml{H.R. Morton and S.G. Lukac, ``The HOMFLY polynomial of the
decorated Hopf link,'' math.GT/0108011.}
\lref\lukac{S.G. Lukac, ``HOMFLY skeins and the Hopf link,'' Ph.D. Thesis,
June 2001, in http://www.liv.ac.uk/~su14/knotgroup.html}
\lref\kz{A. Klemm and E. Zaslow,
``Local mirror symmetry at higher genus,'' hep-th/9906046, in {\it Winter
School on Mirror Symmetry, Vector bundles and Lagrangian
Submanifolds}, p. 183, American Mathematical Society 2001.}
\lref\kkv{S.~Katz, A.~Klemm and C.~Vafa,
``M-theory, topological strings and spinning black holes,'' hep-th/9910181,
Adv.\ Theor.\ Math.\ Phys.\  {\bf 3} (1999) 1445.}
\lref\geomen{S.~Katz, A.~Klemm and C.~Vafa,
``Geometric engineering of quantum field theories,'' hep-th/9609239,
Nucl.\ Phys.\ {\bf B 497} (1997) 173.}
\lref\AVG{M.~Aganagic and C.~Vafa,
``G(2) manifolds, mirror symmetry and geometric engineering,'' hep-th/0110171.}
\lref\ckyz{T.~M.~Chiang, A.~Klemm, S.~T.~Yau and E.~Zaslow,
``Local mirror symmetry: Calculations and interpretations,''
hep-th/9903053,
Adv.\ Theor.\ Math.\ Phys.\  {\bf 3} (1999) 495. }
\lref\proof{H. Ooguri and C. Vafa, ``Worldsheet derivation of a large $N$
duality,'' hep-th/0205297, Nucl.\ Phys.\ {\bf B 641}, 3 (2002).}
\lref\AK{
M.~Aganagic, A.~Karch, D.~Lust and A.~Miemiec,
``Mirror symmetries for brane configurations and branes at singularities,''
Nucl.\ Phys.\ B {\bf 569}, 277 (2000)
[arXiv:hep-th/9903093].}
\lref\AV{M. Aganagic and C. Vafa, ``Mirror symmetry, D-branes and
counting holomorphic discs,'' hep-th/0012041.}
\lref\AKV{M. Aganagic,
A. Klemm and C. Vafa, ``Disk instantons, mirror symmetry and the duality
web,'' hep-th/0105045, Z.\ Naturforsch.\ {\bf A 57} (2002) 1.}
\lref\kl{S. Katz and M. Liu, ``Enumerative geometry of stable
maps with Lagrangian boundary conditions and multiple covers of the disc,''
math.AG/0103074, Adv.\ Theor.\ Math.\ Phys.\  {\bf 5} (2002) 1.}
\lref\rama{P. Ramadevi and T. Sarkar, ``On link invariants and
topological string amplitudes,'' hep-th/0009188, Nucl. Phys. {\bf B 600}
(2001) 487.}
\lref\FI{F.~Cachazo, K.~A.~Intriligator and C.~Vafa,
``A large $N$ duality via a geometric transition,'' hep-th/0103067,
Nucl.\ Phys.\  {\bf B 603} (2001) 3.}
\lref\FII{F.~Cachazo, S.~Katz and C.~Vafa,
``Geometric transitions and N = 1 quiver theories,'' hep-th/0108120.}
\lref\FIII{F.~Cachazo, B.~Fiol, K.~A.~Intriligator, S.~Katz and C.~Vafa,
``A geometric unification of dualities,'' hep-th/0110028,
Nucl.\ Phys.\ {\bf B 628} (2002) 3.}
\lref\ahk{O.~Aharony and A.~Hanany,
``Branes, superpotentials and superconformal fixed points,''
hep-th/9704170, Nucl.\ Phys.\  {\bf B 504} (1997) 239.
O.~Aharony, A.~Hanany and B.~Kol,
``Webs of (p,q) 5-branes, five dimensional field theories and grid
diagrams,'' hep-th/9710116,
JHEP {\bf 9801} (1998) 002.}
\lref\lv{N.~C.~Leung and C.~Vafa,
``Branes and toric geometry,'' hep-th/9711013,
Adv.\ Theor.\ Math.\ Phys.\  {\bf 2} (1998) 91.}
\lref\sft{E.~Witten, ``Noncommutative geometry and string field theory,''
Nucl.\ Phys.\  {\bf B 268} (1986) 253.}
\lref\HV{K.~Hori and C.~Vafa,``Mirror symmetry,'' hep-th/0002222.}
\lref\phases{E.~Witten,``Phases of N = 2 theories in two dimensions,'' hep-th/9301042,
Nucl.\ Phys.\ B {\bf 403}, 159 (1993).}
\lref\kon{M.~Kontsevich, ``Enumeration of rational curves via torus
actions,'' hep-th/9405035, in {\it The moduli space of curves}, p. 335,
Birkh\"auser, 1995.}
\lref\gp{T. Graber and R. Pandharipande, ``Localization of virtual
classes,'' alg-geom/9708001, Invent. Math. {\bf 135} (1999) 487.}
\lref\konts{M. Kontsevich, ``Intersection theory on the moduli space of
curves and the matrix Airy function," Commun. Math. Phys. {\bf 147} (1992)
1.}
\lref\faber{C. Faber, ``Algorithms for computing intersection numbers
of curves, with an application to the class of the locus of Jacobians,"
alg-geom/9706006, in {\it New trends in algebraic geometry},
Cambridge Univ. Press, 1999.}
\lref\cmr{S.~Cordes, G.~W.~Moore and S.~Ramgoolam,
``Lectures on 2-d Yang-Mills theory, equivariant cohomology
and topological field theories,'' hep-th/9411210,
Nucl.\ Phys.\ Proc.\ Suppl.\  {\bf 41} (1995) 184.}
\lref\verlinde{E. Verlinde,
``Fusion rules and modular transformations
in 2-D conformal field theory,''
Nucl.\ Phys.\ {\bf B 300} (1988) 360.}
\lref\BCOV{
M.~Bershadsky, S.~Cecotti, H.~Ooguri and C.~Vafa,
``Kodaira-Spencer theory of gravity and exact results for quantum string
amplitudes,'' hep-th/9309140,
Commun.\ Math.\ Phys.\  {\bf 165} (1994) 311.}
\lref\vaaug{C. Vafa, ``Superstrings and topological strings at large $N$,''
hep-th/0008142, J.\ Math.\ Phys.\  {\bf 42} (2001) 2798.}
\lref\aklm{M.~Aganagic, A.~Karch, D.~Lust and A.~Miemiec,
``Mirror symmetries for brane configurations and branes at singularities,''
hep-th/9903093, Nucl.\ Phys.\ B {\bf 569} (2000) 277.}
\lref\newp{J.M.F.~Labastida and M.~Mari\~no,
``A new point of view in the theory of
knot and link invariants,'' math.QA/0104180, J. Knot Theory
Ramifications {\bf 11} (2002) 173.}
\lref\amv{M.~Aganagic, M.~Mari\~no and C.~Vafa,
``All loop topological string amplitudes from Chern-Simons theory,''
hep-th/0206164.}
\lref\dfg{D.~E.~Diaconescu, B.~Florea and A.~Grassi,
``Geometric transitions and open string instantons,'' hep-th/0205234.
``Geometric transitions, del Pezzo surfaces and open string instantons,''
hep-th/0206163.}
\lref\amer{A. Iqbal, ``All genus topological string amplitudes and 5-brane webs as Feynman  diagrams,'' hep-th/0207114.}
\lref\topstrings{E. Witten, `Topological sigma models,''
Commun.\ Math.\ Phys.\  {\bf 118}, 411 (1988).
``On the structure of the topological phase of two-dimensional gravity,''
Nucl.\ Phys.\ {\bf B 340}, 281 (1990).}
\lref\antibr{C. Vafa, ``Brane/anti-brane systems and $U(N|M)$ supergroup,''
hep-th/0101218.}
\lref\hl{R. Harvey and H.B. Lawson Jr., ``Calibrated geometries,'' Acta Math. {\bf 148} (1982) 47.}
\lref\mayr{P. Mayr, ``Summing up
open string instantons and ${\cal N} = 1$
string amplitudes,'' hep-th/0203237.}
\lref\ik{A.~Iqbal and A.~K.~Kashani-Poor,
``Instanton counting and Chern-Simons theory,'' hep-th/0212279.}
\lref\bmodel{
W.~Lerche and P.~Mayr, ``On ${\cal N} = 1$
mirror symmetry for open type II strings,'' hep-th/0111113.
S.~Govindarajan, T.~Jayaraman and T.~Sarkar,
``Disc instantons in linear sigma models,'' hep-th/0108234,
Nucl.\ Phys.\ B {\bf 646}, 498 (2002).}
\lref\seth{ S.~Sethi, ``Supermanifolds, rigid manifolds and mirror
symmetry,'' Nucl.\ Phys.\ B {\bf 430}, 31 (1994)
[arXiv:hep-th/9404186].
}
\lref\sch{ A.~Schwarz, ``Sigma models having supermanifolds as
target spaces,'' Lett.\ Math.\ Phys.\  {\bf 38}, 91 (1996)
[arXiv:hep-th/9506070].
}
\lref\nv{ A.~Neitzke and C.~Vafa, ``N = 2 Strings and the
Twistorial Calabi-Yau,'' [arXiv:hep-th/0402128].
} \lref\nov{ N. ~Nekrasov, H. ~Ooguri, C.~ Vafa "S-duality and
Topological Strings", arXiv:hep-th/0403167 }
\lref\tsfr{
E.~S.~Fradkin and A.~A.~Tseytlin, ``Conformal Supergravity,''
Phys.\ Rept.\  {\bf 119}, 233 (1985).
}
\lref\tsliu{ H.~Liu and A.~A.~Tseytlin, ``D = 4 super
Yang-Mills, D = 5 Gauged Supergravity, and D = 4 Conformal
Supergravity,''
Nucl.\ Phys.\ B {\bf 533}, 88 (1998) [arXiv:hep-th/9804083].
} \lref\roo{ M.~de Roo and P.~Wagemans, ``Gauge Matter
Coupling In N=4 Supergravity,'' Nucl.\ Phys.\ B {\bf 262}, 644
(1985).
} \lref\mb{N.~ Berkovits and L.~Motl, ``Cubic Twistorial
String Field Theory'', [arXiv:hep-th/0403187].}
\lref\fms{
D.~Friedan, E.~J.~Martinec and S.~H.~Shenker, ``Conformal
Invariance, Supersymmetry And String Theory,'' Nucl.\ Phys.\ B
{\bf 271}, 93 (1986).
}
\lref\hv{ K.~Hori and C.~Vafa, ``Mirror symmetry,''
[arXiv:hep-th/0002222].
}
\lref\wittw{ E.~Witten, ``Perturbative Gauge Theory as a String
Theory in Twistor Space,'' [arXiv:hep-th/0312171].
}
\lref\witol{ E.~Witten, ``An Interpretation Of Classical
Yang-Mills Theory,'' Phys.\ Lett.\ B {\bf 77}, 394 (1978).
}
\lref\agava{ M.~Aganagic and C.~Vafa, ``Perturbative Derivation
of Mirror Symmetry,'' [arXiv:hep-th/0209138].
}

\Title
{\vbox{
 \baselineskip12pt
\hbox{hep-th/0403192}\hbox{HUTP-04/A013}\hbox{UW/PT-04-03}}}
 {\vbox{
 \centerline{Mirror Symmetry and Supermanifolds}
 \centerline{}
 }}
\centerline{Mina Aganagic$^{a}$
and Cumrun Vafa$^{b}$}
\bigskip\centerline{$^{a}$ Department of Physics, University of Washington at Seattle}
\centerline{Seattle, WA 98195-1560, USA}

\bigskip\centerline{$^{b}$ Jefferson Physical Laboratory, Harvard University}
\centerline{Cambridge, MA 02138, USA}\smallskip

\smallskip
 \vskip .3in \centerline{\bf Abstract}

 {We develop techniques for obtaining the mirror
of Calabi-Yau supermanifolds as super Landau-Ginzburg theories.
In some cases the dual can be equivalent to a geometry. We apply
this to some examples.  In particular we show that the mirror of
the twistorial Calabi-Yau ${\bf CP}^{3|4}$ becomes equivalent to a
quadric in ${\bf CP}^{3|3}\times {\bf CP}^{3|3}$ as had been
recently conjectured (in the limit where the K\"ahler parameter of ${\bf CP}^{3|4}$,
 $t\rightarrow \pm
\infty$).  Moreover we show using these techniques that there is
a non-trivial ${\bf Z}_2$ symmetry for the K\"ahler parameter,
$t\rightarrow -t$, which exchanges the opposite helicity states.
As another class of examples, we show that the mirror of certain
weighted projective $(n+1|1)$ superspaces is equivalent to compact
Calabi-Yau hypersurfaces in weighted projective $n$ space.}

 \smallskip \Date{March 2004}

\newsec{Introduction}

Mirror symmetry has been quite effective in clarifying
non-perturbative aspects of the worldsheet theory in
 various contexts (see \ref\mirbook{K. Hori et. al.,
 ``{\it Mirror Symmetry},'', vol. 1 of {\it Clay Mathematics
 Monographs}, AMS, Providence, RI 2003.}).
The basic result of mirror symmetry \hv\
is that the worldsheet
supersymmetric sigma model can be T-dualized to a Landau-Ginzburg
 theory, which in certain cases is equivalent to 
a supersymmetric sigma model
on a different Calabi-Yau manifold.

However, early on it was realized that not every Landau-Ginzburg
theory is equivalent to a Calabi-Yau manifold \ref\cv{C. Vafa,
``Topological Mirrors and Quantum Rings,'' in S.T. Yau (ed.): {\it
Mirror symmetry I}, 121 [arXiv:hep-th/9111017].}. For example,
rigid Calabi-Yau manifolds have a Landau-Ginzburg mirror, but it
cannot have a Calabi-Yau mirror as the mirror manifold cannot have a
K\"ahler class. To overcome this issue it was proposed in \seth\
that one could broaden the space of Calabi-Yau's to include
Calabi-Yau {\it supermanifolds}, to restore the geometric nature
of the mirror symmetry. The mirror of a Calabi-Yau may be a
super-Calabi-Yau. Super-Calabi-Yau manifolds were further studied
in \sch\ where it was shown that as far as A-model topological
strings are concerned certain Calabi-Yau spaces and
super-Calabi-Yau spaces are equivalent.  What was surprising in
this context is that it was found that super-Calabi-Yau's which
are weighted projective superspaces can be equivalent to complete
intersections in ordinary weighted projective spaces.

Our current interest in the subject arose from the conjectures
in \nv .  In that paper the topological
B-model on twistorial Calabi-Yau ${\bf CP}^{3|4}$ \wittw\
was mapped by a conjectured S-duality to A-model on the
same Calabi-Yau and by a further conjectured mirror symmetry
back to a B-model
 on a quadric in ${\bf CP}^{3|3}\times {\bf CP}^{3|3}$.
The motivation for the latter mirror conjecture was that it
was known that the worldsheet instantons are not needed
for a stringy realization of the ${\cal N}=4$ supersymmetric
Yang-Mills on a quadric in ${\bf CP}^{3|3}\times {\bf CP}^{3|3}$,
\witol\ whereas they were expected
to be needed in the A-model on ${\bf CP}^{3|4}$.
Note that these two conjectures
are independent of whether one assumes the relevant
instantans for the gauge sector are closed \wittw\ or open \nv\ Riemann
surfaces.

The S-duality conjecture has now been put on a firmer
ground in \nov\ where it was shown that this topological S-duality
is inherited for ordinary Calabi-Yau threefolds from the S-duality
of type IIB superstrings in 10 dimensions as well as explaining
many other existing results in the literature based on this duality.
In this paper we verify the mirror conjecture of \nv .
More precisely we find that the mirror of ${\bf CP}^{3|4}$
{\it does} depend on the K\"ahler class $t_A$ of ${\bf CP}^{3|4}$.
We show that in the limit $t_A\rightarrow \pm \infty$ the
mirror of ${\bf CP}^{3|4}$ becomes equivalent to a quadric
in ${\bf CP}^{3|3}\times
{\bf CP}^{3|3}$.  We offer an explanation of why $t_A$ enters the
picture:  The value of $t_A$ corresponds
to the expectation value of the dilaton in the ${\cal N}=4$ conformal
supergravity.  It is natural to expect that the observations of \witol\
could be generalized to include the more general geometry
for arbitrary $t$.  It is also conceivable (though we find it less
appealing)
that $t\rightarrow \pm \infty$
is necessary for decoupling the gravity sector from the gauge sector.
At any rate for the Yang-Mills sector by itself the value of $t$
is redundant and can
be reabsorbed into a  redefinition of the gauge coupling constant and
thus can take any value, including $t\rightarrow \pm \infty$, without
affecting the amplitudes.
This thus completes the circle of ideas relating perturbative
Yang-Mills without instantons in the
context of a quadric in ${\bf CP}^{3|3}\times {\bf CP}^{3|3}$
 to its realization in twistor
${\bf CP}^{3|4}$ which requires holomorphic instantons, as is standard
by mirror symmetry.
 
A simple consequence of the above, is the existence of a
${\bf Z}_2$ parity symmetry which exchanges $t_A\leftrightarrow -t_A$.
This symmetry turns out to exchange the two ${\bf CP}^{3|3}$'s and
corresponds to inverting the helicity of the fields.  This is a highly
non-trivial symmetry which was expected from the correspondence
with Yang-Mills theory, but not manifestly realized in the
twistor space ${\bf CP}^{3|4}$.

The method we use to prove the mirror symmetry
is a natural extension of the method in \hv\ to the
case of supermanifolds realized
as linear sigma models with super-chiral fields.  For each
fermionic field we T-dualize the phase of it, which gives
another bosonic mirror field.  However, to conserve
the central charge we also end up with two additional fermions,
i.e., we still have one {\it net} fermionic dimension for the mirror.
This is reminiscent of the bosonization of bosonic ghosts with
first order action \fms .

As another application of our results we provide an alternative
method to deriving the mirror symmetry for compact Calabi-Yau
manifolds.   The results of \sch\ identify the A-model on
Calabi-Yau hypersurfaces with
 A-model on certain weighted projective supermanifolds (without
taking any hypersurface).  It is easier to justify
the mirror symmetry operation in this context because
the circle actions are still symmetries of the theory and
we can dualize them (an alternative strategy was
carried out in \hv ).  We thus use the mirror symmetry
for the {\it non-compact} supermanifold theories with $U(1)$
symmetries to rederive the 
mirror
for ordinary {\it compact} Calabi-Yau manifolds.

The organization of this paper is as follows:
In section 2, after reviewing the standard mirror
symmetry derivation of \hv\ we extend it to the fermionic
case.  In section 3 we give a number of examples
including the twistorial Calabi-Yau and the examples
of  \sch .  In section 4 we discuss
the physical implications of the twistorial mirror.

\newsec{T-duality and a fermionic coordinate}

In this section we define and implement the concept of
 T-duality for a fermionic field.   To do so we first
remind the reader of how it works for the bosonic coordinates \hv .

\subsec{Review of T-duality for bosonic coordinates}

Consider a chiral field $\Phi$ coupled to an $N=2$ $U(1)$
vector multiplet.  Let $\Sigma$ denote the twisted chiral
superfield in the $U(1)$ vector multiplet.  If $\Sigma$
gets a vev, then $\Phi$ picks up mass $Q\Sigma$,
where $Q$ is the $U(1)$ charge of $\Phi$.  The basic
strategy of \hv\ was to first view the $U(1)$ vector
multiplet as a spectator sector and dualize the phase of the
field $\Phi$ thus replacing $\Phi$ by a twisted chiral
field $Y$.  In particular
we define $Y$ by the condition that
$$|\Phi|^2 ={\rm Re} Y$$
and the condition that the phase $\theta$ of $\Phi =|\Phi| e^{i
\theta}$, is T-dual to the shift $\rho$ of $Y+i\rho $ in the
imaginary direction (which is periodic).  As $Y$ and $\Sigma$ are
both twisted chiral fields they can appear in the superpotential
which is a holomorphic function of them
$$W(\Sigma , Y)$$
The form of $W$ is completely fixed by symmetries as follows:
Since $Y$ is periodic it can only appear as a function of
${\rm exp}(-Y)$ except for the fact that $F$, the bottom component of
$\Sigma$, is quantized, and thus allows a term $\Sigma Y$. In fact
this term follows from the action and its coefficient is the
charge of the $\Phi$ field $Q$.  We thus have the term
$$W(\Sigma, Y)=Q\Sigma Y +f({\rm exp}(-Y)).$$
Under T-duality the momentum modes of phase of $\Phi$ are exchanged
with winding modes in the imaginary direction of $Y$. For the BPS
masses to agree, this means that the above LG superpotential
should have critical points shifted in the imaginary part of
$Y$ by $2\pi$.  Moreover the $|\Delta W|$ from one critical
point to the next should be $|Q\Sigma|$ (up
to some normalization conventions), which is the BPS mass
of the momentum mode of $\Phi$.  This uniquely fixes
$f({\rm exp}(-Y))={\rm exp}(-Y)$.  In particular we have
the critical points given by
$$\partial _YW=0\rightarrow {\rm exp}(-Y)=Q\Sigma$$
which are at $Y_0+2\pi i n$ where ${\rm exp}(-Y_0)=Q\Sigma$.
This gives the BPS mass of the winding soliton
$$\Delta W= 2\pi i Q\Sigma$$
 between adjacent
critical points, as was required.  Then the strategy in
\hv\ was to do this for every charge field $\Phi_i$ and
  then recall that $\Sigma$ is also a dynamical
field.   $\Sigma$ couples to the complexified (by the theta angle
of the $U(1)$ gauge theory) FI term which is sometimes denoted by
$t=r+i \theta $.  In other words there is an additional term in
the superpotential
$$W=-t\Sigma$$
(note for example that this leads to the theta angle term for the
$U(1)$ gauge field). Thus, at the level of the superpotential, the
$\Sigma$ appears only in the terms
$$
\Sigma(\sum_i Q_i Y_i-t).
$$
Integrating $\Sigma$ out leads to
\eqn\ycon{\sum_i Q_iY_i =t,}
whose real part is the FI term condition
for the vacuum
$$\sum_i Q_i |\Phi_i|^2 =r.$$
We thus end up with the superpotential
$$W= \sum_i {\rm exp}(-Y_i)$$
where one of the $Y_i$'s can be eliminated using the constraint
\ycon .  This LG theory then gets related to mirror Calabi-Yau geometries.

\subsec{Extension to the super case}

Now we apply the same strategy to the
case of a chiral field $\Theta $ whose top component is fermionic.
We will do this presentation in two steps:  First
following the same steps as the bosonic case. Secondly
we study the effective superpotential for $\Sigma$ and
integrate in the dual fields, along the lines of mirror symmetry
derivation in \agava .

As in the bosonic case, we wish to dualize the phase
of the fermionic field $\Theta$. By this we mean the phase given
by rotating $\Theta$ according to
$$\Theta \rightarrow \Theta \; exp(i\omega).$$
Clearly $\omega$ is bosonic.  So if we dualize it we get a
bosonic dual angle which can now be viewed as the imaginary part
of a twisted chiral multiplet $X$. In particular we set, by a
natural extension of the bosonic case,
$$\Theta \Theta^* =-{\rm Re}X.$$
Moreover the fact that momentum modes
of $\Theta$ should map to winding modes
of $X$ implies that winding sectors
of $X$ should have mass (the choice
of the sign will be explained below) $-Q\Sigma$
in other words we should have
$$W(\Sigma ,X)=-Q\Sigma X +exp(-X).$$
However, introduction of one twisted chiral
bosonic field cannot be the end of the story.  This
theory should have net dimension (i.e.
central charge $\hat c$)  equal to $-1$, whereas $X$ has
net dimension $+1$.  To compensate for this
we should have two additional fermionic fields.  Let
us call them $\eta, \chi$.   This is very much
in the spirit of bosonization of bosonic ghost fields \fms .
 In order for the spectrum
on the two sides to match we note that we start with the
field $\Theta$ which has mass $Q\Sigma$.  The only
way this can be realized in the mirror theory is if
one boson and one fermion cancel in pair from the partition
function. For this we must have $\eta $ and $\chi$ both with mass $Q\Sigma$.
 This
uniquely fixes the superpotential to be
$$W(\Sigma, X, \eta ,\chi)=-Q\Sigma (X-\eta \chi)+ {\rm exp}(-X)$$
Note that the critical points of $W$ with respect to $X,\eta, \chi$
give
$$exp(-X)=-Q\Sigma$$
$$\eta =\chi=0.$$
  At each
of the critical points $X_0+2\pi i n$ with ${\rm exp}(-X_0)=-Q\Sigma$,
the excitations of $\eta $ and $\chi$
 have BPS mass $-Q\Sigma$.  
Also the winding modes of $X$ from one critical point to another
have BPS mass $-Q\Sigma$.  
Thus this gives back the same
net spectrum as the mirror.
Note that we can also shift $X\rightarrow X+\eta \chi$ and
rewrite $W$ as
\eqn\msuppf{
W(\Sigma, X, \eta, \chi)=-Q\Sigma X+{\rm exp}(-X)(1+\eta \chi).
}
Finally, let us explain the sign of $|Q|$ in the superpotential
$W$. This will also provide a consistency check on what we have done
so far.

In the original theory of the fermionic chiral field $\Theta$ of
charge $Q$, integrationg out of $\Theta$ generates an effective
superpotential $W_{eff}(\Sigma)$ for $\Sigma$. This is generated
at one loop, and as fermionic determinants tend to contribute
oppositely to the bosonic determinants, the effective
superpotential is the same as if we were to integrate a bosonic
chiral field of charge $Q$, but of opposite sign: $
W_{eff}(\Sigma) = -Q \Sigma(\log \Sigma -1)$.  The same effective
superpotential must be generated by starting with the dual theory
and integrating out $\eta,\chi$ and $X$. The only effect of
integrating out $\eta$, $\chi$ in \msuppf\ is to change the
measure for $X$, so that $x = e^{-X}$ becomes a good field
variable. Integrating $X$ we recover $W_{eff}(\Sigma)$ with the
correct sign, as claimed.

Again just as in the bosonic case, we do this T-duality for all
charged chiral
fields in the theory, bosonic and fermionic, and then integrate out $\Sigma$, which leaves
us with an effective superpotential
$$
W = \sum_i e^{-Y_i} + \sum_j e^{-X_j}(1+ \eta_j \chi_j)
$$
where $Y_i$'s and $X_j$'s satisfy:
$$
\sum Q_i Y_i -\sum Q_j X_j =t.
$$

\newsec{Examples}
\subsec{Calabi-Yau supermanifolds and hypersurfaces in toric varieties}

In our first application we show that some bosonic Calabi-Yau
manifolds which are hypersurfaces in toric varieties can be
equivalent to super-Calabi-Yau manifolds without hypersurface
constraint! To be precise, the equivalence is strictly speaking
only at the level of F-terms, i.e., at the level of corresponding
topological field theories.  Note that at the level of F-terms we
can replace the fields by their constant modes, in evaluating
periods.  In particular as discussed in \ref\HoriCK{ K.~Hori,
A.~Iqbal and C.~Vafa,
``D-branes and mirror symmetry,''
arXiv:hep-th/0005247.
}\ the periods of a LG theory are evaluated by considering
$$\Pi_\alpha=\int_\alpha \;\prod_i dY_i \;exp(-W),$$
where $\alpha$ denotes a mid-dimensional cycle.
This fact will also be useful for us when we change
variables in order to identify the LG theory with a sigma model geometry.

Calabi-Yau n-fold ${\cal M}$ which is a hypersurface in toric variety is
described
by a theory of
$n+2$ charged chiral fields $X$ and $\Phi_i$, $i=1,\ldots n+1$
with superpotential
$$
\int d^2 \theta \; X P(\Phi_i)
$$
The charges $Q_i$ of $\Phi_i$ and $-Q$ of $X$ satisfy
$$
\sum_i Q_i =Q
$$
and are assumed to be positive $Q, Q_i>0$. For the superpotential
to make sense $P(\Phi_i)$ has to have weight $Q$. This theory has
a phase where it is given by a non-linear sigma model on
Calabi-Yau $Y$ which corresponds to setting
$$
P(\Phi_i)=0
$$
and $X=0$ in
$$
\sum_i Q_i |\Phi_i|^2 - Q|X|^2 =t
$$
modulo gauge transformations.

The claim is that the topological A-model on ${\cal M}$ is the same as the
topological A-model on a super Calabi-Yau manifold $\hat{\cal M}$ with $n+1$
bosonic fields $\Phi_i$ of charge $Q_i$ as above,
but with bosonic chiral field $X$ replaced by a fermionic chiral
 field $\Theta$ of charge $+Q$ and no superpotential.
The corresponding D-term constraint is
$$
\sum_i Q_i |\Phi_i|^2 + Q|\Theta|^2 =t
$$
where the K\"ahler classes of ${\cal M} $ and $\hat {\cal M}$ get
identified.
This equivalence has
 been shown in \sch .  In fact the idea
in \sch\ was to use this equivalence to make the mirror symmetry
of hypersurfaces more manifest, because $\hat {\cal M}$, unlike
${\cal M}$ does have $U(1)$ symmetries which can be
T-dualized\foot{ Note that the compactness of $\cal M$ comes from
imposing $P=0$. While the A-model does not depend on the details
of the superpotential $P(\Phi_i)$, its presence does affect the
A-model theory -- the central charge of the compact theory is
reduced by $2$ with respect to the noncompact case, which is
reflected in the measure factors.  The mirror for $\cal M$ was
constructed from this viewpoint in \hv .}. We will apply our
machinary of mirror symmetry to $\hat {\cal M}$ and show that
indeed it gives rise to the expected mirror B-model of the
corresponding bosonic CY.

The B-model mirror in case of $\hat{\cal M}$ is obtained by simple
application of the above formalism. The path integral is simply
given by
%
$$
\hat{Z} = \int \prod_i dX_i \;dX \; d\eta\; d\chi\; \delta(\sum_i
Q_i X_i - Q X -t)\; exp(\sum_i e^{-X_i} + e^{-X}(1+\eta \chi))
$$
Integrating out $\eta, \chi$ leads just to a measure factor for $X$
$$
\hat{Z} = \int \prod_i dX_i \;e^{-X}dX \;\delta(\sum_i Q_i X_i - Q
X -t) \; exp(\sum_i e^{-X_i} + e^{-X})
$$
Using the delta function constraint we can replace $X$ in $e^{-X}$
term in the measure by $\sum_i{Q_iX_i\over Q} -{t\over Q}$ and ignoring
an irrelevant normalization $e^{-t/Q}$ we have
$$
 \hat{Z} = \int \prod_i dX_i\; e^{-\sum_i Q_iX_i/Q}dX\;\;
\delta(\sum_i Q_i X_i - Q X -t) \;exp(\sum_i e^{-X_i} + e^{-X})
$$
We can use the $\delta $ function to integrate over $X$, which just
fixes its values.  Defining $x_i={\rm exp}(-Q_iX_i/Q)$ (which
requires the standard orbifold procedure on the LG model) we obtain
$$
 \hat{Z} = \int \prod_i dx_i\;
 exp(\sum_i x_i^{Q/Q_i} + e^{t/Q} x_1...x_{n+2})
$$
which is the familiar form of the mirror LG model.
The most familiar case is when $Q_i=1$ and $Q=n+2$.
When $Q$ is not divisible by $Q_i$ we will have to take
further field redefinitions and orbifolds to obtain a suitable
geometric orbifold.
We have thus obtained the mirror for ${\cal M}$, as
was expected.

\subsec{Mirror of ${\bf CP}^{(3|4)}$}

The ${\bf CP}^{(3|4)}$ has a linear sigma model description in
terms of 4 bosonic and 4 fermionic chiral fields
$\Phi^{I}$, $\Theta^I$, $I=0,\ldots 3$ of charges +1. Since
 the sums of the bosonic and fermionic
charges equal, ${\bf CP}^{(3|4)}$ is a super Calabi-Yau manifold.
The lowest components of the bosonic and fermionic fields describe
${\bf C}^{(4|4)}$, and the vacua correspond to
setting D-term potential to zero
$$
\sum_{I=0}^3 |\Phi^I|^2 + \sum_{I=0}^3|\Theta^I|^2 = r
$$
and dividing by the gauge group. The K\"ahler parameter\foot{
For simplicity of notation
here we denote the K\"ahler parameter by $t$ rather than $t_A$.
The S-duality relates it to $t_B/g_B$ \nov\ of the corresponding
B-model.} $r$ is
complexified by the $\theta$ angle to $t=r+i\theta$.

The Landau-Ginzburg mirror of ${\bf CP}^{(3|4)}$ is, as discussed
in the previous section
$$
\int \prod_{I=0}^{3} dY_I\; d X_I \; d\eta_I \;d\chi_I\;
\delta(\sum_{I=0}^3 Y_I - X_I -t) \;exp({\sum_{I=0}^3 e^{-Y_I} +
e^{-X_I}+e^{-X_I}\eta_I \chi_I})
$$
where 
$${\rm Re} Y_I=|\Phi^I|^2$$
$${\rm Re} X_I=-|\Theta^I|^2.$$
We notice that this theory has a ${\bf Z}_2$ symmetry
given by
$$X_I\leftrightarrow Y_I$$
$$\eta_I \rightarrow e^{-X_I} \chi_I$$
$$\chi_I \rightarrow e^{Y_I} \eta_I$$
$$t \rightarrow -t$$
where the change in measure factor for the $\eta_I,\chi_I$ is trivial
thanks to the delta function constraint (up to a trivial
normalization factor).  In other
words, the K\"ahler moduli space of the theory is half the $t$ plane.
This is a highly non-trivial statement as it involves relating
the non-geometric phase $t<0$ to a geometric phase $t>0$.
The meaning of this ${\bf Z}_2$ symmetry will be discussed in the
next section.

We want to rewrite the above as a sigma model on a hypersurface.
Consider putting
$$
X_i = \hat{X}_i + Y_0,\quad Y_i = \hat{Y}_i + Y_0
$$
where $i$ runs as $i=1,2,3$.  After this field redefinition, the
$\delta$ function above effectively sets $X_0$ to
$$
X_0 = \sum_{i=0}^3 ({\hat Y}_i - {\hat X}_i) + Y_0 -t.
$$
Taking all this into account and doing a further
integration over the fermionic fields $\eta_0 ,\chi_0$ gives
us:
$$
\int e^{-Y_0-\sum_i({\hat Y_i}-{\hat X_i})} dY_0\prod_{i=1}^{3}
 d{\hat Y}_i \,d {\hat X}_i \, d\eta_i \, d \chi_i \,exp[e^{-Y_0}(\sum_{i=1}^3 e^{-{\hat Y}_i}
+ e^{-{\hat X}_i} + 1 + e^{t + \sum_{i=0}^3 ({\hat X}_i- {\hat
Y}_i)}+ e^{-{\hat X_i}}\eta_i \chi_i )]
$$
We see that
$$
\Lambda=e^{-Y_0}
$$
 serves as a Lagrange multiplier.
We can introduce ${\bf C}-$valued fields $x_i$ and $y_i$, related
to ${\hat X}_i$, ${\hat Y}_i$ by
$$
e^{-\hat X_i} = x_i , \quad\quad e^{-\hat Y_i} = y_i x_i
$$
and rescale
$$\eta_i\rightarrow e^{\hat{X}_i}\eta_i$$
we find:
$$
\int d\Lambda \int \prod_{i=1}^{3} dy_i\; d x_i d\eta_i d\chi_i
\;\; \exp[\Lambda(\sum_{i=1}^3 x_i y_i + x_i+ 1 + e^{t}
y_1y_2y_3+ \eta_i \chi_i)]
$$
Therefore, the mirror of ${\bf CP}^{(3|4)}$ can be thought of, locally, as
a super-Calabi-Yau hypersurface:
\eqn\ttm{
\sum_{i=1}^3 x_i y_i + x_i+ 1 + e^{t} y_1y_2y_3+ \eta_i \chi_i=0
}
The ${\bf Z_2}$ symmetry which we discussed before
is not apparent in this integrated out version.  However
we can implement it by exchanging the roles of $X_0$ and $Y_0$ as a
Lagrange multiplier, and the field we ``solve'' for
and this leads to
\eqn\sym{
x_i \leftrightarrow y_i, \quad\quad t\leftrightarrow -t
.}
In particular in the other form we end up with the equation
$$\sum_{i=1}^3 x_i y_i + y_i+ 1 + e^{-t} x_1x_2x_3+ \eta_i \chi_i=0.$$
Note that this mirror should still enjoy the original $SU(4|4)$ symmetry
of ${\bf CP}^{3|4}$.  This is not manifest in the above form of the equation,
but it is well known that some symmetries 
becomes less manifest upon T-dualization.  For example
the mirror of supersymmetric sigma model on ${\bf P}^1$ is
an LG theory with
$$W=e^X+e^{-X}$$
and this does not manifestly exhibit the $SU(2)$ symmetry
of ${\bf P}^1$.  The same is true for the mirror of ${\bf CP}^{3|4}$
we have obtained here.

Now we come to the interesting part which motivated this
discussion in the first place.  Namely, we note that in the $ t
\rightarrow -\infty$ limit \ttm\ is local description of a
quadric in ${\bf CP}^{(3|3)} \times {\bf CP}^{(3|3)}$!
That is,
up to a coordinate redefinition \ttm\ becomes: $$ \sum_{i=1}^3 x_i
y_i + 1 + \eta_i \chi_i=0
$$
which is an $x_0 = 1=y_0$
patch of
a quadric
$$
\sum_{I=0}^3 x_I y_I + \eta_i \chi_i=0
$$
in ${\bf CP}^{(3|3)} \times {\bf CP}^{(3|3)}$ with coordinates
$(x^I, \eta^i)$ and $(y^I, \chi^i)$, respectively. Moreover, the
${\bf Z_2}$ symmetry of the parent theory simply becomes the
exchange symmetry of the two ${\bf CP}^{(3|3)}$ factors. We will
discuss the significance of these observations in the context of
twistor space description of ${\cal N} =4 $ SYM in the following
section.

\newsec{Interpretation for twistorial Calabi-Yau}

In the previous section we have studied the B-model mirror of the topological
A-model on ${\bf CP}^{3|4}$.    We have seen that the A-model amplitude
does depend on the K\"ahler structure  parameter $t_A$, as is reflected
by the fact that the B-model mirror's complex structure depends
on $t_A$.   Moreover we have seen that the K\"ahler
modulus has a ${\bf Z_2}$ symmetry:
\eqn\zt{t_A\rightarrow -t_A.}
Furthermore we have shown that in the limit
\eqn\tlim{t_A\rightarrow \pm \infty}
 the mirror geometry
simplifies to a quadric in ${\bf CP}^{3|3}\times {\bf CP}^{3|3}$.

The aim of this section is twofold: First we would like to argue
that the above mirror symmetry in the limit \tlim\ implements the
proposal of \nv\ in relating the two different twistorial
approaches to ${\cal N}=4$ Yang-Mills. Secondly we would
like to give an interpretation of the ${\bf Z}_2$ symmetry \zt .
We will argue that this is equivalent to reversing the role of
positive and negative helicity states, which is a-priori not
manifest in the twistor approach\foot{ For a proposal of how this
is realized in the approach of \ref\berk{N. Berkovits, ``An
alternative string theory in twistor space for ${\cal N}=4$
super-Yang-Mills,'' [arXiv:hep-th/0402045]}\ see the recent paper
\mb .}, but is expected from the viewpoint of the
parity symmetry of the 4 dimensional
theory.

We have seen that the topological strings
on ${\bf CP}^{3|4}$ depends on two parameters:  One is
the string coupling constant $g_A$ and the other is the
K\"ahler class $t_A$ of ${\bf CP}^{3|4}$.  We would thus expect
that the corresponding gravity theory, which is expected to be
${\cal N}=4$ conformal supergravity \wittw\ also has
two free parameters\foot{We have greatly
benefited from discussions with A. Tseytlin in preparation
of this section.}.
Indeed the ${\cal N}=4$ conformal supergravity appears
to have {\it two} free parameters \lref\brw{
E.~Bergshoeff, M.~de Roo and B.~de Wit,
``Extended Conformal Supergravity,''
Nucl.\ Phys.\ B {\bf 182}, 173 (1981).
}
\refs{\brw,\tsfr}:
$$S=\int {1\over \lambda^2} [(C_{\mu\nu\rho\sigma})^2 +S'(\Phi,...)]$$
where $C$ denotes the Weyl tensor, $\Phi$ denotes a complex
scalar field and
$\lambda$ is the coupling constant of the theory.
 In principle we thus have two parameters
we can control independently:  $\lambda$ and $\langle \Phi
\rangle $.  The full structure of the Lagrangian is not
known, but there is no indication that $\lambda$ can be reabsorbed
into the vev of $\Phi$.  In particular if one imposes an $SL(2,R)$
symmetry (whose discrete version is expected to be a true symmetry)
the terms of the form $f(\Phi) C^2$ are ruled out \tsliu .  Moreover the recent connections
with the gauge sector induced conformal gravity in the context of
AdS/CFT suggests that such terms are not generated 
\lref\ts{
A.~A.~Tseytlin,
``On limits of superstring in AdS(5) x S**5,''
Theor.\ Math.\ Phys.\  {\bf 133}, 1376 (2002)
[Teor.\ Mat.\ Fiz.\  {\bf 133}, 69 (2002)]
[arXiv:hep-th/0201112].
}
\refs{\tsliu, \ts}.
This is consistent with the fact that both $g_A$ and $t_A$
deform the topological string theory.
In the context of a stringy realization we would expect $\lambda$
to be identified with the string coupling constant
\eqn\sco{\lambda =g_A}
and
\eqn\dil{\langle \Phi -{\overline \Phi}\rangle =t_A.}
This identification is also natural from the following
viewpoint:  It is known that we need a 2-form field
in addition to the holomorphic form on the ${\bf CP}^{3|4}$
to give the correct gravity fields of the ${\cal N}=4$ conformal
supergravity \ref\witu{E. Witten, private communication.}.  In
\nv\  this field was identified with the K\"ahler class
of ${\bf CP}^{3|4}$.  Thus it is natural to identify
the expectation value for this K\"ahler class with the
expectation value of a field in the gravity sector and
the only choice for this is the expectation value
for the field $\Phi$.   Why does it have to be this
particular identification between $ \Phi  $
and $t_A$? We will now turn to answering this question.

To address this, we need to know how the ${\cal N}=4$
SYM couples to CSG, and in particular
how the $\Phi$ couples to SYM.  This has been studied in
\roo .  See also the recent discussion of this
in \tsliu :
$$S_{YM}=\int {1\over g_A} [{e}^\Phi  \ {\rm tr} F_+\wedge F_+ 
+{e}^{\overline \Phi} \ {\rm tr} F_-\wedge F_- +...]$$
where the $1\over g_A$ in front we obtain
automatically in the string realization of the ${\cal N}=4$ SYM
via D-branes.
There is a symmetry in this theory obtained by the exchange
of the two helicity states:
$$F_+\leftrightarrow F_-\qquad \Phi \leftrightarrow {\overline \Phi}$$
In the context of the twistorial realization of this theory \wittw ,
the instantons are related to the relative strength of these two
terms, i.e.
$${\rm exp}(-t_B/g_B)={\rm exp}({\overline \Phi}-\Phi).$$
Given the S-duality conjecture in \nv , now confirmed in \nov ,
we thus find
$${\rm exp}(-t_A)={\rm exp}({\overline \Phi}-\Phi),$$
which leads us to
$$t_A=\Phi-{\overline \Phi}$$
Thus the helicity exchange symmetry acts as
$$t_A \rightarrow -t_A$$
as was demonstrated in the previous section \zt\ using mirror symmetry!
 As discussed in the previous section this
${\bf Z}_2$ symmetry exchanges the two ${\bf CP}^{3|3}$'s.
However the holomorphic functions on each ${\bf CP}^{3|3}$ signify
helicities of one kind \witol .  More precisely
they correspond to helicity states
$$1(1)+3({1\over 2})+3(0)+1({-1\over 2})$$
$$\qquad \qquad \ \ \ \ \ \ \ \ 1({1\over 2})+3(0)+3({-1\over 2})+1(-1)$$
respectively.  In particular exchanging the two
${\bf CP}^{3|3}$ exchanges the +1 helicity state
with the $-1$ helicity state.  Thus we see that the \zt\
which is also accompanied with exchanging the two ${\bf CP}^{3|3}$'s
exchanges the two opposite helicity states.  This thus gives
a satisfactory explanation of the miraculous fact
that the amplitudes of ${\cal N}=4$ YM do possess this symmetry
which is rather non-trivial in the twistorial formulation
of it.

Now we come to the question of the relation of the
two twistor approaches \refs{\wittw ,\witol},
which was conjectured in \nv\ to be related by S-duality
and a mirror symmetry.  We have
shown they are equivalent only in the limit $t\rightarrow
\pm \infty$.  This is sufficient to establish their equivalence
at the level of the Yang-Mills sector as the $t$ parameter
is redundant in the Yang-Mills sector and can be reabsorbed to 
redefining the gauge coupling constant.  Nevertheless it is natural to ask
what happens if $t$ is finite.
There are two natural scenarios:  One possibility is that the approach 
of
\witol\ could be generalized to any value of $t$.  This we find very
natural because the mirror does make sense for any value
of $t$ and does possess a hidden $SU(4|4)$ symmetry
for any value of $t$ and so it is hard to see what other field theory
one can get on the mirror side.  There is one other logical possibility:
It may be that this story requires
 decoupling of gravity and that for this we need to take 
the $t\rightarrow
\pm \infty$.  This is logically possible
 because  the
gauge sector depends only on the combination $\tau_{YM}=e^{\Phi}/g_A$,
but we find it less appealing.

\bigskip
\noindent {\bf Acknowledgments}
\bigskip
\noindent
We would like to thank N. Berkovits, L.
Motl, A. Neitzke and A. Tseytlin for helpful discussions. The
research of M.A. was supported in part by DOE grant
DE-FG02-96ER40956, by a DOE OJI Award and an Alfred P. Sloan
Foundation fellowship. The research of C.V.~was supported in part
by NSF grants PHY-0244821 and DMS-0244464.
\listrefs
\end
\bye